\documentstyle[12pt]{article}

\skewchar\fivmi='177
\skewchar\sixmi='177
\skewchar\sevmi='177
\skewchar\egtmi='177
\skewchar\ninmi='177
\skewchar\tenmi='177
\skewchar\elvmi='177
\skewchar\twlmi='177
\skewchar\frtnmi='177
\skewchar\svtnmi='177
\skewchar\twtymi='177
\def\@magscale#1{ scaled \magstep #1}
\font\twfvmi  = ammi10   \@magscale5 
\skewchar\twfvmi='177
\skewchar\fivsy='60
\skewchar\sixsy='60
\skewchar\sevsy='60
\skewchar\egtsy='60
\skewchar\ninsy='60
\skewchar\tensy='60
\skewchar\elvsy='60
\skewchar\twlsy='60
\skewchar\frtnsy='60
\skewchar\svtnsy='60
\skewchar\twtysy='60
\font\twfvsy  = amsy10   \@magscale5 
\skewchar\twfvsy='60

\catcode`@=11
\def\un#1{\relax\ifmmode\@@underline#1\else
	$\@@underline{\hbox{#1}}$\relax\fi}
\catcode`@=12

\let\du=\d			
\let\um=\H			

\def\a{\alpha}
\def\b{\beta}

\def\d{\delta}
\def\e{\epsilon}

\def\g{\gamma}

\def\i{\iota}

\def\l{\lambda}

\def\r{\rho}
\def\s{\sigma}

\def\G{\Gamma}

\def\L{\Lambda}

\font\sc=font005			
\def\Sc#1{{\hbox{\sc #1}}}	
\font\ooo=circle10			
\font\ro=manfnt				
\def\kcl{{\hbox{\ro 6}}}		
\def\kcr{{\hbox{\ro 7}}}		
\def\ktl{{\hbox{\ro \char'134}}}	
\def\ktr{{\hbox{\ro \char'135}}}	
\def\kbl{{\hbox{\ro \char'136}}}	
\def\kbr{{\hbox{\ro \char'137}}}	

\def\ip{{=\!\!\! \mid}}                                    
\def\bo{{\raise.15ex\hbox{\large$\Box$}}}		
\def\pr{\prod}						
\def\TH{{\raise.2ex\hbox{$\displaystyle \bigodot$}\mskip-4.7mu \llap H \;}}
\def\face{{\raise.2ex\hbox{$\displaystyle \bigodot$}\mskip-2.2mu \llap {$\ddot
	\smile$}}}					

\def\sp#1{{}^{#1}}				
\def\Tilde#1{{\widetilde{#1}}\hskip 0.03in}			
\def\Hat#1{\widehat{#1}}			
\def\Bar#1{\overline{#1}}			
\def\leftrightarrowfill{$\mathsurround=0pt \mathord\leftarrow \mkern-6mu
	\cleaders\hbox{$\mkern-2mu \mathord- \mkern-2mu$}\hfill
	\mkern-6mu \mathord\rightarrow$}
\def\dvec#1{\vbox{\ialign{##\crcr
	\leftrightarrowfill\crcr\noalign{\kern-1pt\nointerlineskip}
	$\hfil\displaystyle{#1}\hfil$\crcr}}}		
\def\dt#1{{\buildrel {\hbox{\LARGE .}} \over {#1}}}	

\def\frac#1#2{{\textstyle{#1\over\vphantom2\smash{\raise.20ex
	\hbox{$\scriptstyle{#2}$}}}}}			
\def\ha{\frac12}					
\def\sfrac#1#2{{\vphantom1\smash{\lower.5ex\hbox{\small$#1$}}\over
	\vphantom1\smash{\raise.4ex\hbox{\small$#2$}}}}	
\def\bfrac#1#2{{\vphantom1\smash{\lower.5ex\hbox{$#1$}}\over
	\vphantom1\smash{\raise.3ex\hbox{$#2$}}}}	
\def\afrac#1#2{{\vphantom1\smash{\lower.5ex\hbox{$#1$}}\over#2}}    

\newskip\humongous \humongous=0pt plus 1000pt minus 1000pt
\def\caja{\mathsurround=0pt}
\def\eqalign#1{\,\vcenter{\openup2\jot \caja
	\ialign{\strut \hfil$\displaystyle{##}$&$
	\displaystyle{{}##}$\hfil\crcr#1\crcr}}\,}
\newif\ifdtup
\def\panorama{\global\dtuptrue \openup2\jot \caja
	\everycr{\noalign{\ifdtup \global\dtupfalse
	\vskip-\lineskiplimit \vskip\normallineskiplimit
	\else \penalty\interdisplaylinepenalty \fi}}}
\def\li#1{\panorama \tabskip=\humongous				
	\halign to\displaywidth{\hfil$\displaystyle{##}$
	\tabskip=0pt&$\displaystyle{{}##}$\hfil
	\tabskip=\humongous&\llap{$##$}\tabskip=0pt
	\crcr#1\crcr}}

\def\ref#1{$\sp{#1)}$}

\parskip=\medskipamount			
\thicklines			    

\def\oldheadpic{				
	\setlength{\unitlength}{.4mm}
	\thinlines
	\par
	\begin{picture}(349,16)
	\put(325,16){\line(1,0){4}}
	\put(330,16){\line(1,0){4}}
	\put(340,16){\line(1,0){4}}
	\put(335,0){\line(1,0){4}}
	\put(340,0){\line(1,0){4}}
	\put(345,0){\line(1,0){4}}
	\put(329,0){\line(0,1){16}}
	\put(330,0){\line(0,1){16}}
	\put(339,0){\line(0,1){16}}
	\put(340,0){\line(0,1){16}}
	\put(344,0){\line(0,1){16}}
	\put(345,0){\line(0,1){16}}
	\put(329,16){\oval(8,32)[bl]}
	\put(330,16){\oval(8,32)[br]}
	\put(339,0){\oval(8,32)[tl]}
	\put(345,0){\oval(8,32)[tr]}
	\end{picture}
	\par
	\thicklines
	\vskip.2in}
\def\oldtitle#1#2#3#4{\oldheadpic\begin{center}\vglue.5in{\large\bf #1}\\[.6in]
	{#2}\\[.1in] {\it Department of Physics and Astronomy}\\
	{\it University of Maryland, College Park, MD 20742}\\[.6in]
	Physics Publication \#{#3}\\ {#4}\\[1.5in] {\bf Abstract}\\[.1in]
	\end{center} \begin{quotation}}			
\def\oldTitle#1#2#3#4#5#6#7{\oldheadpic\begin{center} \vglue .4in
	{\large\bf #1}\\[.4in]
	{#2}\\[.1in] {\it Department of Physics and Astronomy}\\
	{\it University of Maryland, College Park, MD 20742}\\[.1in]
	{#3}\\[.1in] {\it {#4}}\\ {\it {#5}}\\[.4in]
	Physics Publication \#{#6}\\ {#7}\\[.5in] {\bf Abstract}\\[.1in]
	\end{center} \begin{quotation}}			
\def\border{						
	\setlength{\unitlength}{1mm}
	\newcount\xco
	\newcount\yco
	\xco=-24
	\yco=12
	\begin{picture}(140,0)
	\put(\xco,\yco){$\ktl$}
	\advance\yco by-1
	{\loop
	\put(\xco,\yco){$\kcl$}
	\advance\yco by-2
	\ifnum\yco>-240
	\repeat
	\put(\xco,\yco){$\kbl$}}
	\xco=158
	\yco=12
	\put(\xco,\yco){$\ktr$}
	\advance\yco by-1
	{\loop
	\put(\xco,\yco){$\kcr$}
	\advance\yco by-2
	\ifnum\yco>-240
	\repeat
	\put(\xco,\yco){$\kbr$}}
        \put(-20,11){\tiny University of Maryland Elementary Particle
Physics University of Maryland Elementary Particle Physics University of
Maryland Elementary Particle Physics}
	\put(-20,-241.5){\tiny University of Maryland Elementary
Particle Physics University of Maryland Elementary Particle Physics
University of Maryland Elementary Particle Physics}
	\end{picture}
	\par\vskip-8mm}
\def\bordero{						
	\setlength{\unitlength}{1mm}
	\newcount\xco
	\newcount\yco
	\xco=-24
	\yco=12
	\begin{picture}(140,0)
	\put(\xco,\yco){$\ktl$}
	\advance\yco by-1
	{\loop
	\put(\xco,\yco){$\kcl$}
	\advance\yco by-2
	\ifnum\yco>-240
	\repeat
	\put(\xco,\yco){$\kbl$}}
	\xco=158
	\yco=12
	\put(\xco,\yco){$\ktr$}
	\advance\yco by-1
	{\loop
	\put(\xco,\yco){$\kcr$}
	\advance\yco by-2
	\ifnum\yco>-240
	\repeat
	\put(\xco,\yco){$\kbr$}}
	\put(-20,12){\ooo
bacdefghidfghghdhededbihdgdfdfhhdheidhdhebaaahjhhdahbahgdedgehgfdiehhgdigicba}
	\put(-20,-241.5){\ooo
ababaighefdbfghgeahgdfgafagihdidihiidhiagfedhadbfdecdcdfagdcbhaddhbgfchbgfdacfediacbabab}
	\end{picture}
	\par\vskip-8mm}
\def\headpic{						
	\indent
	\setlength{\unitlength}{.4mm}
	\thinlines
	\par
	\begin{picture}(29,16)
	\put(165,16){\line(1,0){4}}
	\put(170,16){\line(1,0){4}}
	\put(180,16){\line(1,0){4}}
	\put(175,0){\line(1,0){4}}
	\put(180,0){\line(1,0){4}}
	\put(185,0){\line(1,0){4}}
	\put(169,0){\line(0,1){16}}
	\put(170,0){\line(0,1){16}}
	\put(179,0){\line(0,1){16}}
	\put(180,0){\line(0,1){16}}
	\put(184,0){\line(0,1){16}}
	\put(185,0){\line(0,1){16}}
	\put(169,16){\oval(8,32)[bl]}
	\put(170,16){\oval(8,32)[br]}
	\put(179,0){\oval(8,32)[tl]}
	\put(185,0){\oval(8,32)[tr]}
	\end{picture}
	\par\vskip-6.5mm
	\thicklines}
\def\title#1#2#3#4{\border\headpic {\hbox to\hsize{#4 \hfill UMDEPP #3}}\par
	\begin{center} \vglue .5in {\large\bf #1}\\[.6in]
	{#2}\\[.1in] {\it Department of Physics and Astronomy}\\
	{\it University of Maryland, College Park, MD 20742}\\[1.5in]
	{\bf Abstract}\\[.1in] \end{center} \begin{quotation}}	
\def\Title#1#2#3#4#5#6#7{\border\headpic
	{\hbox to\hsize{#7 \hfill UMDEPP #6}}\par
	\begin{center} \vglue .4in {\large\bf #1}\\[.4in]
	{#2}\\[.1in] {\it Department of Physics and Astronomy}\\
	{\it University of Maryland, College Park, MD 20742}\\[.1in]
	{#3}\\[.1in] {\it {#4}}\\ {\it {#5}}\\[.5in] {\bf Abstract}\\[.1in]
	\end{center} \begin{quotation}}			
\def\endtitle{\end{quotation}\newpage}			

\def\sect#1{\bigskip\medskip \goodbreak \noindent{\bf {#1}} \nobreak \medskip}
\def\refs{\sect{References} \footnotesize \frenchspacing \parskip=0pt}
\def\Item{\par\hang\textindent}

\begin{document}

\def\scst{\scriptstyle}
\def\itrema{$\ddot{\scriptstyle 1}$}
\def\Bo{\bo{\hskip 0.03in}}
\def\lrad#1{ \left( A {\buildrel\leftrightarrow\over D}_{#1} B\right) }
\def\derx{\partial_x} \def\dery{\partial_y} \def\dert{\partial_t}
\def\Vec#1{{\overrightarrow{#1}}}
\def\.{.$\,$}

\def\grg#1#2#3{Gen.~Rel.~Grav.~{\bf{#1}} (19{#2}) {#3} }
\def\pla#1#2#3{Phys.~Lett.~{\bf A{#1}} (19{#2}) {#3}}

\def\ula{{\underline a}} \def\ulb{{\underline b}} \def\ulc{{\underline c}}
\def\uld{{\underline d}} \def\ule{{\underline e}} \def\ulf{{\underline f}}
\def\ulg{{\underline g}} \def\ulm{{\underline m}}
\def\uln#1{\underline{#1}}
\def\ulp{{\underline p}} \def\ulq{{\underline q}} \def\ulr{{\underline r}}

\def\hatm{\hat m}\def\hatn{\hat n}\def\hatr{\hat r}\def\hats{\hat s}
\def\hatt{\hat t}

\def\plpl{{+\!\!\!\!\!{\hskip 0.009in}{\raise -1.0pt\hbox{$_+$}}
{\hskip 0.0008in}}}

\def\mimi{{-\!\!\!\!\!{\hskip 0.009in}{\raise -1.0pt\hbox{$_-$}}
{\hskip 0.0008in}}}

\def\items#1{\\ \item{[#1]}}
\def\ul{\underline}
\def\un{\underline}
\def\-{{\hskip 1.5pt}\hbox{-}}

\def\kd#1#2{\d\du{#1}{#2}}
\def\fracmm#1#2{{{#1}\over{#2}}}
\def\footnotew#1{\footnote{\hsize=6.5in {#1}}}

\def\low#1{{\raise -3pt\hbox{${\hskip 1.0pt}\!_{#1}$}}}

\def\ip{{=\!\!\! \mid}}
\def\unb{{\underline {\bar n}}}
\def\upb{{\underline {\bar p}}}
\def\um{{\underline m}}
\def\up{{\underline p}}
\def\Phib{{\Bar \Phi}}
\def\Phit{{\tilde \Phi}}
\def\Phibt{{\tilde {\Bar \Phi}}}
\def\Db{{\Bar D}_{+}}
\def\gg{{\hbox{\sc g}}}
\def\nt{$~N=2$~}

\vskip 0.07in

{\hbox to\hsize{June 1993\hfill UMDEPP 93--213}}\par

\begin{center}
\vglue .35in

{\large\bf Self--Dual ~Supersymmetric ~Yang--Mills ~Theory ~Generates}\\
\vskip 0.02in
{\large\bf Two--Dimensional ~Supersymmetric ~WZNW ~Models}$\,$\footnote{This
work is supported in part by NSF grant \# PHY-91-19746.} \\[.1in]

\baselineskip 10pt

\vskip 0.30in

Hitoshi ~NISHINO\footnote{E-Mail (internet): Nishino@UMDHEP.umd.edu} \\[.2in]
{\it Department of Physics} \\ [.015in]
{\it University of Maryland at College Park}\\ [.015in]
{\it College Park, MD 20742-4111, USA} \\[.1in]
and\\[.1in]
{\it Department of Physics} \\[.015in]
{\it Howard University} \\[.015in]
{\it Washington, D.C. 20059, USA} \\[.18in]

\vskip 1.5in

{\bf Abstract}\\[.1in]
\end{center}

\begin{quotation}

{}~~~We show that recently formulated four-dimensional self-dual
supersymmetric Yang-Mills theory, which is consistent background
for open $~N=2$~ superstring, generates two-dimensional
$~N=(1,1),~\, N=(1,0) $~ and $~N=(2,0)$~ supersymmetric gauged
Wess-Zumino-Novikov-Witten $~\s\-$models on coset manifolds $~G/H$, after
appropriate dimensional reductions.  This is supporting evidence for the
conjecture that the self-dual supersymmetric Yang-Mills
theory will generate lower-dimensional supersymmetric integrable models after
dimensional reductions.

\endtitle

\def\doit#1#2{\ifcase#1\or#2\fi}
\def\[{\lfloor{\hskip 0.35pt}\!\!\!\lceil}
\def\]{\rfloor{\hskip 0.35pt}\!\!\!\rceil}
\def\delsl{{{\partial\!\!\! /}}}
\def\caldsl{{\calD\!\!\! /}}
\def\calO{{\cal O}}
\def\asym{({\scriptstyle 1\leftrightarrow \scriptstyle 2})}
\def\Lag{{\cal L}}
\def\du#1#2{_{#1}{}^{#2}}
\def\ud#1#2{^{#1}{}_{#2}}
\def\dud#1#2#3{_{#1}{}^{#2}{}_{#3}}
\def\udu#1#2#3{^{#1}{}_{#2}{}^{#3}}
\def\calD{{\cal D}}
\def\calM{{\cal M}}
\def\tildef{{\tilde f}}
\def\calDsl{{\calD\!\!\!\! /}}

\def\Hat#1{{#1}{\large\raise-0.02pt\hbox{$\!\hskip0.038in\!\!\!\hat{~}$}}}
\def\hati{{\hat{I}}}
\def\dt{$~D=10$~}
\def\alp{\alpha{\hskip 0.007in}'}
\def\oalp#1{\alp^{\hskip 0.007in {#1}}}
\def\naive{{{na${\scriptstyle 1}\!{\dot{}}\!{\dot{}}\,\,$ve}}}
\def\items#1{\vskip 0.05in\Item{[{#1}]}}
\def\item#1{\Item{#1}}

\def\pl#1#2#3{Phys.~Lett.~{\bf {#1}B} (19{#2}) #3}
\def\np#1#2#3{Nucl.~Phys.~{\bf B{#1}} (19{#2}) #3}
\def\prl#1#2#3{Phys.~Rev.~Lett.~{\bf #1} (19{#2}) #3}
\def\pr#1#2#3{Phys.~Rev.~{\bf D{#1}} (19{#2}) #3}
\def\cqg#1#2#3{Class.~and Quant.~Gr.~{\bf {#1}} (19{#2}) #3}
\def\cmp#1#2#3{Comm.~Math.~Phys.~{\bf {#1}} (19{#2}) #3}
\def\jmp#1#2#3{Jour.~Math.~Phys.~{\bf {#1}} (19{#2}) #3}
\def\ap#1#2#3{Ann.~of Phys.~{\bf {#1}} (19{#2}) #3}
\def\prep#1#2#3{Phys.~Rep.~{\bf {#1}C} (19{#2}) #3}
\def\ptp#1#2#3{Prog.~Theor.~Phys.~{\bf {#1}} (19{#2}) #3}
\def\ijmp#1#2#3{Int.~Jour.~Mod.~Phys.~{\bf {#1}} (19{#2}) #3}
\def\nc#1#2#3{Nuovo Cim.~{\bf {#1}} (19{#2}) #3}
\def\ibid#1#2#3{{\it ibid.}~{\bf {#1}} (19{#2}) #3}

\def\szet{{${\scriptstyle \b}$}}
\def\ula{{\un a}}
\def\ulb{{\un b}}
\def\ulc{{\un c}}
\def\uld{{\un d}}
\def\ulA{{\un A}}
\def\ulM{{\underline M}}
\def\cdm{{\Sc D}_{--}}
\def\cdp{{\Sc D}_{++}}
\def\vTheta{\check\Theta}
\def\Pisl{{\Pi\!\!\!\! /}}

\def\fracmm#1#2{{{#1}\over{#2}}}
\def\gg{{\hbox{\sc g}}}
\def\half{{\fracm12}}
\def\ha{\half}

\def\frac#1#2{{\textstyle{#1\over\vphantom2\smash{\raise -.20ex
	\hbox{$\scriptstyle{#2}$}}}}}			

\def\fracm#1#2{\hbox{\large{${\frac{{#1}}{{#2}}}$}}}

\def\Dot#1{\buildrel{_{_{\hskip 0.01in}\bullet}}\over{#1}}
\def\dt#1{\Dot{#1}}
\def\uln{{\underline n}}
\def\Tilde#1{{\widetilde{#1}}\hskip 0.015in}
\def\Hat#1{\widehat{#1}}

\def\Dot#1{\buildrel{_{_{\hskip 0.01in}\bullet}}\over{#1}}
\def\dt#1{\Dot{#1}}
\def\gg{{\hbox{\sc g}}}
\def\nt{$~N=2$~}
\def\gg{{\hbox{\sc g}}}
\def\nt{$~N=2$~}
\def\tr{{\rm tr}}
\def\Tr{{\rm Tr}}
\def\mpl#1#2#3{Mod.~Phys.~Lett.~{\bf A{#1}} (19{#2}) #3}
\def\hati{{\hat i}} \def\hatj{{\hat j}} \def\hatk{{\hat k}}
\def\hatl{{\hat l}}

\oddsidemargin=0.03in
\evensidemargin=0.01in
\hsize=6.5in
\textwidth=6.5in

\noindent 1.~~{\it Introduction.~~~}There has been recently considerable
development related to $~N=2$~ superstring theories [1].  Among others,
the most interesting aspect is that four-dimensional $(D=4)$~ self-dual
$~N=4$~ supersymmetric Yang-Mills (SDSYM) theory is a consistent background
for the {\it open} $~N=2$~ superstring [2].  Originally, self-dual Yang-Mills
(SDYM) theory has very interesting feature related to a fascinating
mathematical conjecture [3] that {\it all} (bosonic) integrable models in
lower-dimensions are generated by the $~D=4$~ SDYM theory.  It is then natural
trial to generalize this conjecture to supersymmetric case, namely
we can expect the SDSYM theory will generate
{\it supersymmetric} integrable models in lower-dimensions.  It has been
also noticed recently that {\it closed} $~N=2$~ superstring has
$~N=8$~ self-dual supergravity (SDSG) theory [2] as its consistent
backgrounds.

	Based on the above conjecture, we have recently studied various
aspects of SDSYM and SDSG theories [4], and we have shown many examples
of embedding supersymmetric integrable models into $~D=4$~ SDSYM theory,
{\it e.g.,} the supersymmetric KdV systems [5],
supersymmetric $~SL(n)$~ Toda gauge theory together with $~W_\infty\-$gravity
as its $~n\rightarrow \infty$~ limit [6],
supersymmetric KP hierarchy [7], and Witten's topological field
theory in $~D=2$~ [6].  It has been also found that there
exist some exact solutions for the SDSG system [8], indicating possibilities
of other integrable models.

	An interesting question then is whether or not the same SDSYM
theory can also generate lower-dimensional (especially $~D=2$)
supersymmetric Wess-Zumino-Novikov-Witten (WZNW) $~\s\-$models on coset
manifold $~G/H$~ for arbitrary Lie gauge groups $~G$~ and $~H$.  In
this Letter we answer this question for the first time by showing the
explicit examples of $~N=(1,1),~\, ~N=(1,0)$~ and $~N=(2,0)$~
supersymmetric WZNW models in $~D=2$, embedded into the $~D=4$~ SDSYM theory.

\bigskip\bigskip

\noindent 2.~~{\it Embedding of $~N=(1,1)$~ WZNW Model.~~~}As the first
simple example, we deal with the $~D=2,~N=(1,1)$~ supersymmetric WZNW model [9]
embedded into the $~D=4$~ SDSYM [2,4].  To this end, we start with a
resulting set of superfield equations of the latter system after appropriate
dimensional reductions and truncations.  Needless to say, this method of
going down from $~D=4$~ theory to $~D=2$~ system is equivalent to the other
way around of embedding the latter into the former.

	Among many possible dimensional reductions of the SDSYM theory,
the easiest choice is the simple dimensional reduction [10] with
appropriate truncations.  Skipping
the details which have been explained in our previous paper [5], we
recall that the resultant superspace field equation in $~D=2$~ is simply the
vanishing of superfield strength for a Yang-Mills gauge
group:
$$ F\du{A B}I \equiv D_A A_B{}^I - (-)^{A B} D_B A_A {}^I
- T\du{A B} C A\du C I + f^{I J K} A_A{}^J A_B{}{}^K = 0~~.
\eqno(2.1) $$
Here $~f^{I J K}$~ is the structure constant for the gauge group $~G$, and
the adjoint indices $~{\scst I,~J,~\cdots}$~ will be usually omitted below.
Generally, the range of indices $~{\scst A,~B,~\cdots}$~ depends on the
number of
supersymmetries $~N$.  For example, for the present choice of $~N=(1,1)$~
we have $~{\scst (A) ~=~ (\a,~ a)~ =~ (+,~-,~\plpl,~\mimi)}$, where
$~{\scst +,~-}$~ are chiral spinorial coordinates, while the last two are for
the light-cone coordinates.  The equation $~F_{A B}=0$~ is the result of the
component self-duality condition $~F=\Tilde F$~ after our
truncation applied to the $~N=4$~ SDSYM [2,4] as the consistent background
for the $~N=2$~ superstring, or to its descendant $~N=2$~
SDSYM theory [5] after appropriate truncation.  In any case, eq.~(2.1) is
eventually the simplest
choice of the truncation applied to the $~N=4$~ SDSYM [4].  (See also
Ref.~[6].)  Accordingly, in this dimensional reduction scheme [5],
the maximal value of $~N$~ is $~N=(4,4)$.

	We next review the system of $~N=(1,1)$~ gauged WZNW model on $~G/H$~
[9].  Its total action is
$$\eqalign{S (g, L_A, R_A) =
& \, - \half \int d^4 z ~\left[ \Tr \left( \Hat \Pi^\a \Hat \Pi_\a
\right) +2 \int_0 ^1 d y ~ (\g_5)^{\a\b}~ \Tr \left(
\Tilde \Pi_\a \Tilde\Pi_\b \Tilde \Pi_y  \right) \right] \cr
& + i \int d^4 z ~ (\g_5)^{\a\b} ~\Tr \left( L_\a g
D\low\b g^{-1} - R_\a g^{-1} D\low\b g
-i g R_\a g^{-1} L_\b \right)~~. \cr }
\eqno(2.2) $$
Here $~\int d^4 z \equiv \int d^2 x d^2 \theta$~ is the usual
$~N=(1,1)$~ superspace integral, while
$$\Pi_A \equiv g^{-1} D\low A g ~~,~~~~\Hat \Pi_A \equiv \Pi_A + i g^{-1}
L\low A g - i R_A \equiv g^{-1} \nabla\low A g ~~,
\eqno(2.3) $$
and ~$(D_A)\equiv (D_\a, \partial_a)$, where
the $~D=2$~ supergravity backgrounds are omitted for simplicity.
The $~g$~ is a gauge group-valued scalar superfield: $~g \equiv
\exp(\varphi^I T^I)$~ with
the generators $~T^I$~ of $~G$, and an appropriate scalars $~\varphi^I$.
The $~L_A$~ and $~R_A$~ are the $~H\-$gauge
superfields, which are gauge group-valued on the subgroup
$~H_L\otimes H_R$.
All the {\it tilded} quantities are functions of $~(x^m,y)$~ defined
{\it e.g.}~by $~\Tilde\Pi _A(x,y=1) = \Pi_A(x),~ \Tilde\Pi _A(x,y=0) = 0$,
and $~\Tilde\Pi_y \equiv \Tilde g^{-1} \partial_y \Tilde g$~ [9].
The total action is invariant under the local $~H\-$gauge transformation:
$$\eqalign{&\d g = i g \L_R - i \L_L g~~, \cr
&\d L_A = D_A \L_L + i \[ L_A, \L_L \]~~, ~~~~
\d R_A = D_A \L_R + i \[ R_A, \L_R \] ~~, \cr }
\eqno(2.4)$$
iff only the {\it diagonal} subgroup $~H_D$~ of $~H_L\otimes H_R$~ is
gauged [9]:
$$L_A = R_A ~~, ~~~~\L_L = \L_R ~~.
\eqno(2.5) $$

	The most relevant to our purpose are the field equations,
which are
$$\eqalign{&\left( 2\Hat \Pi _- - i g^{-1} L_- g \right) \bigg| _H = 0 ~~,
{}~~~~L_+ |_H = 0 ~~, \cr
&\left( 2\Hat \Pi _+  + i R_+ \right) \bigg| _H = 0 ~~,
{}~~~~R_-|_H = 0 ~~, \cr
&\nabla_- \Hat\Pi _+ - 2 R_{- +} = 0 ~~, ~~~~ R_{+ -} = 0 ~~, \cr }
\eqno(2.6) $$
where [9]
$$\eqalign{&\nabla_\a \Hat\Pi_\b \equiv D_\a \Hat\Pi_\b
+ i \{ R_\a , \Hat\Pi_\b \}~~, \cr
& 2i R_{\a\b} \equiv D_\a R_\b + D_\b R_\a + i \{ R_\a,R_\b\} ~~, \cr}
\eqno(2.7) $$
and the symbol $~|_H$~ denotes the restriction of the quantity on the
subgroup $\, H$.  Now combining these equations, we clearly see that
$$\li{ & L_\a = 0 ~~, ~~~~R_\a = 0 ~~,
&(2.8a) \cr
&\Pi_+ \big| _H = 0 ~~, ~~~~\Pi_-\big| _H = 0 ~~, \cr
&D_- \Hat \Pi_+ = D_- \Pi_+ = 0 ~~,
&(2.8b) \cr} $$
also due to (2.5).  Here we no longer have to distinguish $~\Hat\Pi_A$~ from
$~\Pi_A$~ due to (2.8a).

	Our ansatz for embedding this system into (2.1) is as follows:
We identify the superpotentials $~A_A$~ in the latter with
$$\eqalign{ & A_- = 0 ~~, ~~~~ A_{\mimi}= 0 ~~, \cr
& A_+ = \Pi_+ ~~, ~~~~ A_{\plpl} = \Pi_{\plpl} ~~.  \cr }
\eqno(2.9) $$
Most of the components
of $~F_{A B}$~ vanish automatically, but the non-trivial ones which
need the use of superfield equation (2.8) are
$$\eqalign{ &F_{+ - } = D_- \Pi_+ = 0 ~~, \cr
& F_{\plpl - } = - D_- \Pi_\plpl = 0 ~~, ~~~~
F_{\mimi +} = 2D_- (D_- \Pi_+) = 0 ~~, \cr
& F_{\plpl\mimi} = - 2 D_- ( D_- \Pi_\plpl ) = 0 ~~.\cr}
\eqno(2.10) $$
Here $~D_- \Pi_\plpl= 0$~ is shown by using the relationship
$$\eqalign{0 &= D_+ (D_- \Pi_+) = - D_- D_+ (g^{-1} D_+ g)  \cr
& = (D_- \Pi_+) \Pi_+ - \Pi_+ (D_- \Pi_+) - \fracm i 2 D_- \Pi_\plpl ~~.
\cr}
\eqno(2.11) $$

	Interesting feature here is that we have identified what is
usually called {\it current} superfields $~\Pi_A$~ with the {\it potential}
superfields $~A_A$~ in our {\it new} superspace.\footnotew{To avoid
confusion, we are using $~A_A$~ instead of $~\G_A$~ usually used in
the original superspace.  The superpotential $~A_A$~ is {\it not}
to be confused with the independent superfields $~L_A$~ and $~R_A$,
either.}~~In
this sense, our new superspace is to be {\it distinguished} from the
original superspace, where the composite gauge fields are defined.
Another point to be stressed is the special role
played by the {\it on-shell} condition (2.8a), which made the
distinction between $~\Pi_A$~ and $~\Hat\Pi_A$~ no longer important,
once the superfield equations are used.  In other words,
our special identification of the usual current superfields with
superpotential is possible only {\it on-shell}.

\bigskip\bigskip

\noindent 3.~~{\it Geometrical Interpretation.~~~}We can
interpret our identification (2.9) in a more geometrical and convincing
way.  Notice that a $~\s\-$model on a group manifold $~G$~ can be also
described by a metric formulation, where we use the coordinates
$~(\phi^i)~~{\scst(i~=~1,~2,~\cdots,~{\rm dim}\,G)}$~, the metric
$~g\low{i j}$~
and the vielbein $~e\low I{}^i$~ of $~G$.  As usual $~g\low{i j} = e\du i I
e\du j I$, {\it etc}.  In this formulation, the bosonic part of the
$~\s\-$model (2.2) before gauging takes a somewhat familiar
form:
$$S_{\rm bosons} = \int d^2 x~ \bigg[ \half g\low{i j}
(\phi) (\partial_m \phi^i) (\partial ^m \phi^j)
+ \int_0^1 d y~ f^{I J K} \e^{m n} \Tilde e_{i I}\Tilde e_{j J}
\Tilde e_{k K} (\partial_m \Tilde\phi^i)  (\partial_n \Tilde\phi^j)
(\partial_y \Tilde\phi^k)  \bigg] ~~.
\eqno(3.1) $$
The meaning of {\it tilded} fields is the same as in (2.2).
The translation rule between our original group element notation and
this metric formulation is illustrated by\footnotew{This relationship has been
also presented in Ref.~[11].}
$$ g^{-1} \partial_m g = (\partial_m \phi^i) e\du i I T^I ~~,
\eqno(3.2) $$
or its superspace generalization
$$ g^{-1} \partial{\low M} g = (\partial_M \Phi^i) e\du i I T^I ~~,
\eqno(3.3) $$
where $~\Phi^i$~ is now a scalar superfield as a function of the
appropriate superspace coordinates $~(z^M)$, containing $~\phi^i$~ as
its $~\theta=0$~ sector.  We can easily confirm the validity of (3.3) for
the $~N=(1,1)$~ superspace, by rewriting the original action
(2.2) for the {\it ungauged} case in terms of
$~\phi^i$~ and $~e\du i
I$~ or $~g\low{i j}$.  One of the important relations obtained from (3.3)
is
$$F\du{i j} I \equiv \partial_i e\du j I - \partial_j e\du i I
+ f^{I J K} e\du i J e\du j K = 0 ~~.
\eqno(3.4) $$
This is derived by applying a differential operator on the
differential 1-form $~g^{-1} d g \equiv (d\Phi^i)e\du i I T_I \equiv (d
Z^M) (\partial_M \Phi^i) e\du iI T^I \equiv e ^I T_I\equiv e$,
as follows:
$$d(g^{-1} d g ) = ( de^I) T_I \equiv d e ~~,
\eqno(3.5) $$
where the l.h.s.~is
$$d(g^{-1} d g ) = (d g^{-1} )\wedge d g = - g^{-1} (d g) \wedge g ^{-1} d g
= - e\wedge e~~.
\eqno(3.6) $$
Equating (3.5) with (3.6), we get (3.4) as
$$ 0 = d e + e \wedge e = - (d Z^M) (\partial_M \Phi^i) \wedge (d Z^N)
(\partial_N \Phi^j) F\du{i j} I T^I \equiv - (d Z^N)\wedge (d Z^M)
F\du{M N} I T^I ~~,
\eqno(3.7) $$
also yielding (2.1).\footnotew{We can reconfirm this in terms of tensor
component calculation.}~~Eq.~(3.4) and (3.7) imply that the vielbein
$~e\du i I$~ on
$~G$~ can be identified with a {\it gauge} field $~A\du i I$~ for $~G$, and
moreover its pull-back onto $~D=2$~ base superspace gives the composite gauge
field:
$$A\du M I T_I \equiv (\partial _M \Phi^i) A\du i I T_I = (\partial_M
\Phi^i) e\du i I T_I = g^{-1} \partial{\low M} g~~,
\eqno(3.8) $$
which is nothing else than a generalization of our identification (2.9),
before the gauging.
The $~\theta=0~$ sector of (3.7) has been already known as the
structure equation for the Lie group manifold $~G$~ [11], and our
relations are just its superspace generalization.

	From this geometrical consideration, it is natural to expect
that our identification is universal for any supersymmetric $~\s\-$model
on group manifold.  In the previous $~N=(1,1)$~ case, since $~L_A$~ and
$~R_A$~ gauge superfields vanish {\it on-shell}, the validity of above
identification
is intact even after the gauging by $~L_A$~ and $~R_A$.  This enables us to
generalize the target manifold from $~G$~ to a more general coset manifold
$~G/H$.  It has also provided natural justification of identifying the
usual ``current superfields'' with ``potential''
superfields.

\bigskip\bigskip

\noindent 4.~~{\it Embedding of $~N=(1,0)$~ WZNW Model.~~~}We now
see similar embeddings is possible for the $~N=(1,0)$~
non-chiral and chiral WZNW models developed
by Gates {\it et al.} [12].  Since these systems are similar to the previous
$~N=(1,1)$~ case which we are already used to, and moreover we have
understood the general geometrical treatments, we give here mainly the
superfield equations with some minor explanations about notations.  The
only subtlety is due to the unidexterous feature of the systems, which
needs detailed study even though we know it should work also here from
our geometrical consideration.\footnotew{For the metric formulation for the
$~\s\,$-models with unidexterous supersymmetries, see Ref.~[13].}

	The total action of the {\it non-chiral} $~N=(1,0)$~ WZNW model
on $~G/H$~ contains the superfields $~g,~\G_+$~ and
$~\G_\mimi$:
$$\eqalign{kS(g,\G_+,\G_\mimi) = &- \fracmm{i k}{2\pi}
\int d^3 z ~\Tr \left[ \Pi_+ \Pi_\mimi
+ \int_0^1 d y ~ \Tilde \Pi_y \left( \Tilde \Pi _+ \Tilde\Pi_\mimi -
\Tilde\Pi _\mimi \Tilde\Pi_+ \right) \right] \cr
& + \fracmm{i k} \pi \int d^3 z~ \Tr \left[( \partial_\mimi g) g ^{-1} \G_+ -
\G_\mimi g^{-1} D_+ g + g^{-1} \G_+ g \G_\mimi - \G_+ \G _\mimi \right]
{}~~,
\cr}
\eqno(4.1) $$
where $~\Pi$'s are defined as in (2.3) with the ~$G\-$valued field $~g$.
The $~\G_+, ~\G_\mimi$~ and $~g$~ superfield equations are
$$\li{& \left( g^{-1} \nabla_+ g\right) \big|_H = 0 ~~, ~~~~
[ ( \nabla_\mimi g) g^{-1} ] |_H = 0 ~~,
&(4.2a) \cr
&\nabla_\mimi (g^{-1} \nabla_+g) = \partial_\mimi\G_+ - D_+\G_\mimi + \[
\G_+, \G_\mimi \] \equiv W_-(\G) ~~,
&(4.2b) \cr} $$
where $~\nabla_A$~ are the $~H\-$gauge covariantization by $~\G_+$~ and
$~\G_\mimi$, namely $~\nabla_A ~ * \equiv D_A - \[ \G_A, * \]$.
Combining (4.2a) and (4.2b), we see that the $~H\-$field strength
vanishes: $~W_- (\G) =0$~ {\it on-shell}, and also owing to the $~H\-$gauge
invariance of the action, we can choose a special gauge
where $~\G_+$~ and $~\G_\mimi$~ are zero.  This enables us to get rid of
the effective presence of $~\G_A$~ and eventually our superfield
equations are
$$\eqalign{ &(g^{-1} D_+ g )|_H = [ (\partial_\mimi g) g^{-1}
]\big| _H = 0 ~~, \cr
& \partial_\mimi (g^{-1} D_+ g) = 0 ~~, ~~~~ D_+ \[ ( \partial _\mimi g)
g^{-1} ] = 0 ~~. \cr }
\eqno(4.3) $$

	We can now embed these superfield equations into the $~N=(1,0)$~
version of our master equation (2.1) for the components $~{\scst A ~=~
+, ~\plpl, ~\mimi}$~ by the following identifications:
$$\eqalign{& A_+ = \Pi_+ ~~, ~~~~ A_\plpl = \Pi_\plpl ~~, \cr
& A_ \mimi = 0 ~~, ~~~~(A_+ |_H = 0) ~~.  \cr }
\eqno(4.4) $$
Now all the components $~F_{+ +}, ~F_{+ \plpl}, ~ F_{+\mimi},
{}~F_{\plpl\mimi}$~ vanish,\footnotew{Notice that components such as $~F_{+
-}$~ do {\it not} exist in the $~N=(1,0)$~ superspace.} and the only
non-trivial ones which need the superfield equation (4.3) are
$$F_{+\mimi} = - \partial_\mimi (g^{-1} D_+ g) = 0 ~~,
{}~~~~ F_{\plpl\mimi} = - \partial_\mimi ( g^{-1} \partial_\plpl g ) = 0 ~~,
\eqno(4.5) $$
where the last one vanishes for the same reason as in (2.11).
We have thus confirmed that our universal rule can be also applied to
the $~N=(1,0)$~ non-chiral WZNW model.

	We finally mention the embedding of $~N=(1,0)$~ {\it chiral}
WZNW model on $~G/H$~ with {\it chiral bosons} [12].  This model is also called
$~G/H$~ lefton model, in which the $~G\-$valued $~g\-$fields are called chiral
bosons $~g\low L$, and moving only into the left
direction.  For this to be realized {\it on-shell}, the system requires
a Lagrange multiplier superfield $~\L\du + \mimi$.  The total action is
$$\eqalign{ k_L S(g\low L, \G_+, \G_\mimi, \L\du +\mimi) =
{}~&- \fracmm{ik_L} {2\pi} \int d^3 z ~ \Tr \left[ \Pi^L_+ \Pi^L_\mimi + \L\du
+
\mimi \Pi^L_\mimi \Pi^L_\mimi + \int_0^1 d y ~ \Tilde \Pi^L_y( D_+
\Tilde \Pi^L_\mimi -\partial_\mimi \Tilde \Pi^L_+) \right]  \cr
& + \fracmm{ik_L}\pi \int d^3 z~ \Tr \bigg[ (\partial_\mimi g\low L)
g_L^{-1} \G_+ - \G_\mimi g_L^{-1} (D_+ g\low L + \L\du +
\mimi \partial_\mimi g\low L ) \cr
& ~~~~~ ~~~~~ ~~~~~ ~~~~~ ~~~~~ + g_L^{-1} \G_+ g\low L \G_\mimi
- \G_+ \G_\mimi\bigg] ~~, \cr}
\eqno(4.6) $$
where $~\Pi^L_A\equiv g_L^{-1} D_A g\low L, ~\Tilde\Pi^L_y\equiv
\Tilde g_L^{-1} \partial_y \Tilde g\low L$,
and $~\L\du + \mimi$~ is a Lagrange
multiplier.  Leaving all other details to Ref.~[12],
we give the superfield equations:
$$\eqalign{& (g_L^{-1} D_+ g\low L ) \big| _H =
\left[ (\partial_\mimi g\low L) g_L^{-1} \right]\bigg| _H = 0 ~~, \cr
& \partial_\mimi \left[ g^{-1} \low L D_+ g\low L  \right] = 0 ~~, ~~~~
\Tr \left[ (\partial_\mimi g\low L) g_L^{-1} \right]^2 = 0 ~~. \cr }
\eqno(4.7) $$
Here we have already chosen a special gauge $~\G_A = 0, ~\L\du +
\mimi=0$~ [12], so that the original $~H\-$gauge covariant derivatives
$~\nabla_A$~ are now ordinary $D_A$'s.

	The embedding of this system now follows the same pattern as
(4.4), namely
$$\eqalign{& A_+ = g_L^{-1} D_+ g\low L ~~, ~~~~ A_\plpl = g_L^{-1} \partial
_\plpl g\low L ~~, \cr
& A_- = 0 ~~, ~~~~ A_ \mimi = 0 ~~, ~~~~(A_+ |_H = 0) ~~.  \cr }
\eqno(4.8) $$
Needless to say, here again the {\it on-shell} triviality of
$~\G_A$~ made the embedding possible.
We can get similar results for $~G/H$~ righton model, which we skip
here.

\bigskip\bigskip

\noindent 5.~~{\it Embedding of $~N=(2,0)$~ WZNW Model.~~~}We finally
show the embedding of the $~N=(2,0)$~ non-chiral gauged WZNW models, which
are more important for realistic model building with the target space-time
supersymmetry based on the $~D=4,\, N=1$~ superstring [14].  In the
algebraic approach, we can formulate the coset models using (minimal) conformal
field theories [15].  The corresponding component lagrangian formulation with
$~N=(2,2)$~ conformal supersymmetry has been also recently presented [16].
In this Letter we present for simplicity the $~N=(2,0)$~ case, which
can be obtained by some truncation of the former.  The truncation can be
easily done, if we delete what is called $~\chi\-$field in Ref.~[16],
together with the right-handed supersymmetries in $~N=(2,2)$.

	Due to the $~N=(2,0)$~ superconformal invariance of the
$~\s\-$model, the the gauge
groups $~G$~ and $~H$~ can be no longer arbitrary here, and the coset is
required to be K{\" a}hler manifold [13].  Moreover, to comply with the
operator product expansion of the $~N=2$~ superconformal Kac-Moody
algebra [15], the manifold is further restricted to be
hermitian symmetric space [17], which is a special kind of K\" ahler
manifold.  A typical example is $~G=SU(n+m),~H=SU(n)\otimes
SU(m)\otimes U(1)$~ [15], and the coset is what is called
complex Grassmannian manifold.  The Cartan-Weyl bases for the algebra
of $~G$~ are $~H_{\hat i}~{\scst({\hat i}~=~ 1,~2,~\cdots,~{\rm rank}~G)}$,
$~E_\i$~ and
$~E^\i$, where $~{\scst \i}$~ is for the positive roots of the algebra
of $~G$.  Similarly the coset $~G/H$~ can be spanned by the bases $~E_i$~
and $~E^i$~ with the positive roots $~{\scst i}$~ in the algebra of
$~G$~ but not in that of $~H$.

	The total action for our $~N=(2,0)$~ gauged WZNW model
obtained by the truncation above is
$$\eqalign{ S(g,\G_\plpl,\, &\G_\mimi,\l_{+1}, \l_{+2}) \cr
= \,&-\fracm k{4\pi} \int d^2 x ~ \Tr\Bigg[ g^{-1} (\partial_\plpl g) g^{-1}
(\partial_\mimi g) \cr
& ~~~~~ ~~~~~ ~~~~~ ~~~~~ - \int_0^1 d y~ \Tilde g^{-1} (\partial_y
\Tilde g) \left\{\Tilde g^{-1} (\partial_\plpl \Tilde g)
\Tilde g^{-1} (\partial_\mimi \Tilde g) - \Tilde g^{-1}
(\partial_\mimi\Tilde g) \Tilde g^{-1} (\partial_\plpl\Tilde g)
\right\} \Bigg]  \cr
& + \fracm k {2\pi} \int d^2 x ~\Tr \left[ \G_\plpl(\partial_\mimi g)
g^{-1} - \G_\mimi g^{-1} (\partial_\plpl g) + \G_\plpl g \G_\mimi g^{-1}
- \G_\plpl\G_\mimi \right] \cr
& - \fracm {i k}{2\pi} \int d^2 x ~\l_{+2i} \left[ \partial_\mimi
\l\du{+1} i - (\r^{\hat I} )\du j i \G\du\mimi {\hat I}\l\du{+1} j
\right]~~,  \cr }
\eqno(5.1) $$
in component fields.
The constant matrix $~\r^{\hat I}$~ is
defined by $~\[ T^{\hat I}, E_j \] = (\r^{\hat I})\du j k E_k$.
Here $~T^{\hat I}~{\scst (\hat I ~=~1,~2,~\cdots,~{\rm dim}\, H)}$~ are
the generators of $~H$.  The $~\l\du{+1} i $~ and
$~\l_{+2 i}$~ are chiral Majorana-Weyl fermions.
Accordingly, we get the component field equations, which will be also
dictated as the superfield equations:
$$\li{&\Tr \left[T^{\hat I} g^{-1} \nabla_\plpl g\right]
- i (\r^{\hat I})\du i j \l\du {+1} i \l_{+2 j} = 0 ~~,
&(5.2) \cr
&\Tr \left[ T^{\hat I} (\nabla_\mimi g) g^{-1} \right]  = 0 ~~,
&(5.3) \cr
&\nabla_\mimi (g^{-1} \nabla_\plpl g) = \partial_\mimi \G_\plpl -
\partial_\plpl \G_\mimi + \[ \G_\plpl, \G_\mimi\] \equiv
W_{\mimi\plpl}(\G) ~~,
&(5.4) \cr
&\nabla_\mimi \l\du{+1}i = \nabla_\mimi\l_{+2i} = 0 ~~,
&(5.5) \cr } $$
where $~\nabla_a ~{\scst (\,a ~=~
\plpl,~\mimi)}$~ are $~H\-$gauge covariant derivatives with the
connection $~\G_a \equiv \G\du a {\hat I} T^{\hat I}$
Like the cases of $~N\le 1$~, we can easily
see the {\it on-shell} vanishing of the $~H\-$gauge field strength:
$$ W_{\plpl\mimi} (\G) \equiv \partial_\plpl\G_\mimi - \partial_\mimi \G_\plpl
- \[ \G_\plpl,\G_\mimi\] = 0 ~~,
\eqno(5.6) $$
by comparing (5.4) with (5.2), using also (5.5).

	We can reformulate these in $~N=(2,0)$~ superspace by setting up
the structure equations (supertranslations):
$$\eqalign{&\nabla_{+1} g = - \l\du {+1} i ~g E_i ~~, ~~~~ \nabla_{+2} g =
- \l_{+2 i } ~ g E^i ~~, \cr
&\nabla_{+1} \l_{+2 i } = - i (g^{-1} \nabla_\plpl g)\low i ~~, ~~~~
\nabla_{+2} \l_{+1}{}^i = - i (g^{-1} \nabla_\plpl g)^i ~~, \cr }
\eqno(5.7) $$
where our $~N=(2,0)$~ superalgebra is
$$\{ D_{+1}, D_{+2} \} = i \partial_\plpl ~~,
\eqno(5.8) $$
and all other (anti)commutators among
$~D_A~{\scst(A~=~\plpl,~\mimi,~+1,~+2)}$~
are zero.  Accordingly, the component field equations (5.2) - (5.5) can be
rewritten in superspace easily.  The invariance of (5.1) under (5.7) is
to be performed in terms of ``1.5$\-$order'' formalism {\it without} the
variation with respect to $~\G_\plpl~$ or $~\G_\mimi$.

	We now embed this system into (2.1) now with $~{\scst A
{}~=~\plpl,~\mimi,~ +1,~+2} $~ of $~N=(2,0)$~ superspace.  Following the
results for $~N=(1,0)$~ as a guiding principle,
we choose a particular gauge, in which the superfield $~\G_A$~ has
been gauged away, and set up the ans{\" a}tze:
$$\eqalign{ &A_{+1} = g^{-1} D_{+1} g ~~, ~~~~ A_{+2} = g^{-1} D_{+2} g
{}~~, \cr
&A_\plpl = g^{-1} \partial_\plpl g ~~, ~~~~ A_{\mimi} = 0 ~~,
{}~~~~ (A_{+1} |_H = A_{+2}|_H = 0) ~~. \cr }
\eqno(5.9) $$
We can easily confirm that $~F_{A B} = 0$, by the use of superfield
equations.  Most of these components vanish identically under (5.7), and
all of them vanish after the use of superfield equations and (5.6).
Non-trivial ones are, {\it e.g.},
$$ F_{\mimi +1} = (\partial_\mimi\l\du{+1} i )
E_i = 0 ~~, ~~~~
F_{\plpl\mimi} = - \partial_\mimi \left[ g^{-1} (\partial_\plpl g) \right]
= 0 ~~.
\eqno(5.10) $$

The success of the embedding of the $~N=(2,0)$~ WZNW is an exciting
result, because we know that these models with
$~D=(1,3)\,$\footnotew{The expression $~D=(t,s)$~ implies
$~(t+s)\,$-dimensional space-time with $~t$~ time coordinates and $~s$~
spacial coordinates.}  target
space-time correspond to realistic $~N=1$~ superstring models [14,15].
We have thus realized that the $~N=2$~ superstring is actually
the {\it underlying} theory also for $~D=(1,3),\, N=1$~
superstring models [14,15] {\it via} the $~D=(2,2)$~ SDSYM theory.

\bigskip\bigskip

\noindent 6.~~{\it Concluding Remarks.~~~}In this Letter we have shown
that supersymmetric WZNW models on coset manifolds $~G/H$~ are embedded into
the $~D=4$~ SDSYM theory {\it via} appropriate dimensional reductions,
and truncations.
In other words, the latter theory can generate the former systems as its
descendant theories, as supporting evidence for the conjecture that the
$~D=4$~ SDSYM theory is the ``master theory'' of lower-dimensional
supersymmetric integrable models.

	To our knowledge, our result is the first one to embed
supersymmetric WZNW models into $~D=4$~ SDSYM theory.  We have also
established a guiding principle of identifying the pull-back of vielbein
of the group manifold onto the base $~D=2$~ superspace with a new
superpotential, which satisfies the integrability condition (2.1).
Motivated by the embedding into $~D=4$~ SDSYM theory, we have developed
a new viewpoint for the coset-type WZNW-model in terms of such a
topological condition as (2.1) in superspace.

	The explicit examples given are the $~N=(1,1),~N=(1,0)$~ and
$~N=(2,0)$~ supersymmetric WZNW models on coset manifolds $~G/H$.
The gauging of the subgroup $~H$~ of $~G$~ is used in order to generalize
the $~\s\-$model to the coset $~G/H$.  The most interesting case for
model building is the $~N=(2,0)$~ WZNW models with hermitian symmetric
coset spaces, which correspond to $~D=4,\,N=1$~
superstring theory [14,15].
The key observation is the supersymmetric generalization of the
structure equation (3.4) for the Lie gauge group $~G$, which seems to
be universal to any supersymmetric WZNW $~\s\-$model, and moreover the
introductions of
gauged subgroup $~H$~ does {\it not} invalidate the condition (3.4) at
least after the use of field equations.  This enabled us to generalize
our original results for $~G$~ to a more general coset $~G/H$.

	The results we have obtained in this Letter have also provided strong
evidence supporting the concept that $~N=1$~ superstring models
themselves are the descendant effective theories of $~N=2$~ superstring,
because the former can be described by appropriate coset models [12,14,15].
This philosophy of ``world-sheet for world-sheet'' has been suggested in
various contexts in our paper [6] as well as in Ref.~[18].

	Even though our results here are rather expected from the
above-mentioned conjecture, we can utilize this feature for the understanding
of general deformations of superconformal theories [19].  In particular, the
{\it non-perturbative} aspect of superstring or superconformal theories may
well be understood in terms of the ``master theory'', connecting all
the integrable models as their deformations.
In this Letter we considered only the case
with vanishing field strength as the solution to the $~D=4$~ self-duality
condition, but this is not the most general case, and even non-vanishing field
strengths are possible.  We expect much more topologically
quantized non-trivial backgrounds solutions $~F_{A B} \neq 0$~ connected by
some tunneling between them.   Such quantum effect is likely to give
non-perturbative significance on WZNW models.

	The lagrangian approach we
are using for integrable models seems to be powerful enough to describe
general deformations of superconformal theories [19].  Actually, as has been
studied in Refs.~[12,19], the lagrangian approach provides a wide class of
superconformal coset models.

	According to a recent paper [20], an $~N=2$~ gauged WZNW coset
model of $~SU(2,{\bf R})/U(1)\,\,\,\,\,$
\hbox{is closely related to a two-dimensional
black hole solution [21]} near its singularity.  Based on this result,
we can imagine that the $~D=4$~ SDSYM theory is controlling not only the
coset-type models or other integrable models on {\it flat} backgrounds, but
also other gravitational systems, which are also integrable in
lower-dimensions.

	We mention that the guiding principle in section 3 can
be applied also to $~D=3$~ supersymmetric $~\s\-$models on group
manifolds or cosets.  Eventually we will have the same integrability
condition (2.1) formulated in $~D=3$~ superspace closely related to
supersymmetric Chern-Simons theory [22], which is again an effective
theory obtained from the $~D=4$~ SDSYM.  In other words, the $~D=4$~
SDSYM theory covers such a large set of $~\s\-$models in $~D=3$~ as
well as in $~D=2$, as its descendant theories.  Our results in this
Letter that supersymmetric WZNW coset models are embedded into the
$~N=2$~ superstring theory {\it via} $~D=4$~ SDSYM, combined with the
other examples of embedding integral models, such as supersymmetric KdV
models [5], super Liouville and supersymmetric Toda theory [6], or
supersymmetric KP systems [7], cover a wider class of integrable
systems in $~D= 2$~ and $~D=3$.

\bigskip\bigskip

We are indebted to S.J\.Gates, Jr.~and O.A.~Solovi{\" e}v for helpful
discussions related to supersymmetric WZNW models.

\bigskip\bigskip

\vfill\eject

\refs
\small

\items{1} H.~Ooguri and C.~Vafa, \mpl{5}{90}{1389};
\np{361}{91}{469}; \ibid{367}{91}{83};
\item{  } H.~Nishino and S.J.~Gates, Jr., \mpl{7}{92}{2543}.

\items{2} W.~Siegel, Stony Brook preprint, ITP-SB-92-31 (July, 1992).

\items{3} M.F.~Atiyah, unpublished;
\item{  } R.S\.Ward, Phil.~Trans.~Roy.~Lond.~{\bf A315} (1985) 451;
\item{  } N.J\.Hitchin, Proc.~Lond.~Math.~Soc.~{\bf 55} (1987) 59.

\items{4} S.V.~Ketov, S.J.~Gates, Jr.~and H.~Nishino,
\pl{307}{93}{323};
\item{  } H.~Nishino, S.J.~Gates, Jr. and S.V.~Ketov,
\pl{307}{93}{331};
\item{  } S.J.~Gates, Jr., H.~Nishino and S.V.~Ketov,
\pl{297}{92}{99};
\item{  } S.J.~Ketov, H.~Nishino and S.J.~Gates, Jr., \np{393}{93}{149}.

\items{5} S.J.~Gates, Jr.~and H.~Nishino, \pl{299}{93}{255}.

\items{6} H.~Nishino, Maryland preprint, UMDEPP 93--144, to
appear in Phys.~Lett.~B.

\items{7} H.~Nishino, Maryland preprint, UMDEPP 93-145 (Feb.~1993).

\items{8} H.~Nishino, \pl{307}{93}{339}.

\items{9} P.~Di Vecchia, V.G.~Knizhnik, J.L.~Pertersen and P.~Ross,
\np{253}{85}{701};
\item{  } G.~Atkinson, U.~Chattopadhyay and S.J.~Gates, \ap{168}{86}{387}.

\items{10} J.~Scherk and J.H.~Schwarz, \np{153}{79}{61}.

\items{11} E.~Braaten, T.L.~Curtright and C.K.~Zachos,
\np{260}{85}{630}.

\items{12} S.J.~Gates, Jr., S.V.~Ketov, S.M.~Kuzenko and O.A.~Soloviev,
\np{362}{91}{199}.

\items{13} E.~Bergshoeff, H.~Nishino and E.~Sezgin, \pl{166}{86}{141}.

\items{14} D.~Gepner, \pl{199}{87}{380}, \np{296}{88}{757}.

\items{15} Y.~Kazama and H.~Suzuki, \pl{216}{89}{112}; \np{321}{89}{232}.

\items{16} T.~Nakatsu, \ptp{87}{92}{795}.

\items{17} S.~Helgason, {\it ``Differential Geometry, Lie Groups and
Symmetric Spaces''}, (Academic Press, NY, 1978);
\item{   }  J.A.~Wolf, {\it ``Symmetric Spaces''}, (Marcel Dekkar,
NY, 1972).

\items{18} M.B.~Green, \pl{193}{87}{439}; \np{293}{87}{593};
\item{   } H.~Nishino, \mpl{7}{92}{1805}.

\items{19} S.J.~Gates, Jr.~and O.A.~Solovi{\" e}v, \pl{294}{92}{342};
\item{   } O.A.~Solovi{\" e}v, \mpl{8}{93}{301}.

\items{20} T.~Eguchi, \mpl{7}{92}{85}.

\items{21} E.~Witten, \pr{44}{91}{314};
\item{  } G.~Mandal, A.M.~Sengupta and S.R.~Wadia,
\mpl{6}{91}{1685}.

\items{22} S.J.~Gates, M.T.~Grisaru, M.~Ro{\v c}ek and W.~Siegel, ``{\it
Superspace}'', (Benjamin/Cummings, Reading MA, 1983), page~27;
\item{   } H.~Nishino and S.J.~Gates, Maryland preprint, UMDEPP 92--060,
Int.~Jour.~Mod.~Phys.~{\bf 8A} (1993), to appear.

\end{document}